\documentclass[10pt,conference]{IEEEtran} 
\usepackage{cite}
\usepackage{amsmath,amssymb,amsfonts}
\usepackage{algorithmic}
\usepackage{graphicx}
\usepackage{textcomp}
\usepackage{xcolor}
\usepackage{tcolorbox}
\usepackage{comment}
\usepackage{tabularx}
\usepackage{makecell}
\usepackage{graphicx}
\usepackage{cellspace}
\usepackage{mdframed}
\usepackage{array}
\usepackage{makecell}
\usepackage{balance}
\usepackage[raggedrightboxes]{ragged2e}

\mdfdefinestyle{MyFrame}{
    innertopmargin=4pt,
    innerbottommargin=4pt,
    innerrightmargin=4pt,
    innerleftmargin=4pt,
    backgroundcolor=gray!15!white
        }

\usepackage{xcolor}
 \usepackage{colortbl}
\usepackage{ifthen}
\newboolean{commentEnabled}
\setboolean{commentEnabled}{true}
\newcommand{\zibran}[1]{\ifthenelse{\boolean{commentEnabled}}{\textcolor{red}{Zibran: #1}}{}}
\newcommand{\rabbi}[1]{\ifthenelse{\boolean{commentEnabled}}{\textcolor{blue}{Rabbi: #1}}{}}
\newcommand{\champa}[1]{\ifthenelse{\boolean{commentEnabled}}{\textcolor{pink}{Champa: #1}}{}}

\definecolor{grayish1}{rgb}{0.94902, 0.94902, 0.94902}  %
\definecolor{grayish}{rgb}{0.90196, 0.90196, 0.90196}
\definecolor{yellowish2}{rgb}{1.00000, 1.00000, 0.80000}
\definecolor{yellowish1}{rgb}{1.00000, 1.00000, 0.96078}
\definecolor{yellowish}{rgb}{1.00000, 1.00000, 0.89804}
\definecolor{bluish2}{rgb}{0.80000, 0.80000, 1.00000}
\definecolor{bluish1}{rgb}{0.89804, 0.89804, 1.00000}
\definecolor{bluish0}{rgb}{0.96078, 0.96078, 1.00000}
\definecolor{bluish}{rgb}{0.92549, 0.92549, 0.97255}
\definecolor{greenish1}{rgb}{0.89804, 1.00000, 0.89804}
\definecolor{greenish}{rgb}{0.96078, 1.00000, 0.96078}		
\definecolor{pinkish2}{rgb}{0.92157, 0.69804, 0.69804}
\definecolor{pinkish1}{rgb}{1.0000 , 0.8000 , 0.8000}		
\definecolor{pinkish}{rgb}{1.0000 , 0.8780  , 0.8780 }
\definecolor{pink}{rgb}{1.0000 , 0.7 , 0.7 }

\usepackage{todonotes}

 \title{Chasing the Clock: How Fast Are Vulnerabilities Fixed in the Maven Ecosystem?}

\author{\IEEEauthorblockN{
Md Fazle Rabbi ~~~~~~~~ Arifa Islam Champa ~~~~~~~~ Rajshakhar Paul ~~~~~~~~ Minhaz F. Zibran}
\IEEEauthorblockA{\textit{Department of Computer Science, Idaho State University, Pocatello, ID, United States} \\
\{mdfazlerabbi, arifaislamchampa, rajshakharpaul, zibran\}@isu.edu}
}

\begin{document}

\maketitle

\begin{abstract}
This study investigates the software vulnerability resolution time in the Maven ecosystem, focusing on the influence of CVE severity, library popularity as measured by the number of dependents, and version release frequency. The results suggest that critical vulnerabilities are addressed slightly faster compared to lower-severity ones. 
Library popularity shows a positive impact on resolution times, while frequent version updates are associated with faster vulnerability fixes. 
These statistically significant findings are based on a thorough evaluation of over 14 million versions from 658,078 libraries using the dependency graph database of Goblin framework. These results emphasize the need for proactive maintenance strategies to improve vulnerability management in open-source ecosystems.
\end{abstract}

\begin{IEEEkeywords}
Goblin, dependency, vulnerability, vulnerability fixing, library popularity, library release frequency 
\end{IEEEkeywords}

\section{Introduction}


Software vulnerabilities continue to pose significant threats to the security and reliability of modern software systems. 
Open-source libraries are integrated into millions of applications which enable collaboration~\cite{zhang2020companies} and innovation~\cite{meszaros2024dynamics} on a global scale.  
However, this widespread use of open-source software also brings security challenges. A recent report revealed that 84\% of analyzed codebases contained at least one known vulnerability, with 48\% classified as high or critical severity~\cite{bals2024open}. These statistics signify the need for effective vulnerability management.



Despite ongoing efforts to secure software ecosystems ~\cite{alfadel2023empirical, wang2023deepvd, zhang2023mitigating, okafor2022sok}, identification, prioritization, and resolution of vulnerabilities in open-source libraries remain challenging.
Severe vulnerabilities, which should ideally be addressed immediately, often face delays due to resource limitations~\cite{ptsecurity2024consequences}. 
Timely resolution of software vulnerabilities is crucial for reducing security risks and maintaining system integrity~\cite{sen2020determinants}. 
However, vulnerabilities often go unaddressed by many organizations. Tenable reports that vulnerabilities from 2017 are still being exploited, and most zero-day vulnerabilities exploited in 2022 were disclosed on same day patches were released~\cite{venkat2023unpatched}. 
This highlights a persistent delay in addressing vulnerabilities promptly.



Popular libraries are expected to fix vulnerabilities quickly due to active communities, but the high volume of issues can overwhelm maintainers and cause delays~\cite{brewer2021know}. 
Maintenance activity adds another layer of complexity to this issue. 
Many open-source projects lack regular maintenance which leave them vulnerable to prolonged security issues~\cite{weigel2024silent}.
The absence of timely updates can expose the project to security risks, quickly diminishing trust among users and contributors~\cite{weigel2024silent}. Understanding how factors such as severity, popularity, and maintenance activity interact to influence fix times is crucial for improving vulnerability management practices. 

In this study, we investigate the factors influencing the time required to fix vulnerabilities in software libraries within the Maven ecosystem. Specifically, we examine whether severity levels of common vulnerabilities and exposures (CVEs), number of dependent versions, and version release frequency of libraries affect the speed of vulnerability resolution. To explore these factors, we address the following three research questions (RQs):

\vspace{2pt}
\noindent \textbf{RQ-1:} \textit{How do average fix times for vulnerabilities compare across different CVE severity levels?}

\noindent
-- Investigating how the severity of CVE affects vulnerability fixing time will reveal how vulnerabilities are prioritized. 
This will help developers create better strategies and allocate resources to enhance security while reducing exploitation risks. Moreover, this will provide the research community with insights into industry standards and effective vulnerability management.

\vspace{2pt}
\noindent \textbf{RQ-2:} \textit{Do more popular libraries resolve vulnerabilities faster than less popular ones?}

\noindent
-- By determining whether the popularity of software libraries influences the speed at which their vulnerabilities are resolved, we can understand the dynamics of resource allocation and community responsiveness within the open-source ecosystem. These insights will guide library selection decisions which will balance popularity with security responsiveness.

\vspace{2pt}
\noindent\textbf{RQ-3:} \textit{Do actively maintained libraries fix vulnerabilities faster than those that are less maintained?}

\noindent
-- Assessing the effect of the release frequency of software libraries on the time it takes to resolve vulnerability issues is important for ensuring timely security updates and reducing the risk of prolonged vulnerability exposure. 

\vspace{2pt}
To address the above three research questions, we follow the procedural steps shown in Figure~\ref{fig:method} and analyze over 14 million versions of 6,58,078 libraries using Goblin framework~\cite{Jaime2025navigating}. 
To ensure reproducibility, the datasets as well as the scripts for data preprocessing, statistical analysis, and visualizations used in this study are made publicly available~\cite{replicationPackage}.

\section{Dataset}\label{sec:data}
\begin{figure*}[htbp]
\centerline{\includegraphics[scale=0.57]{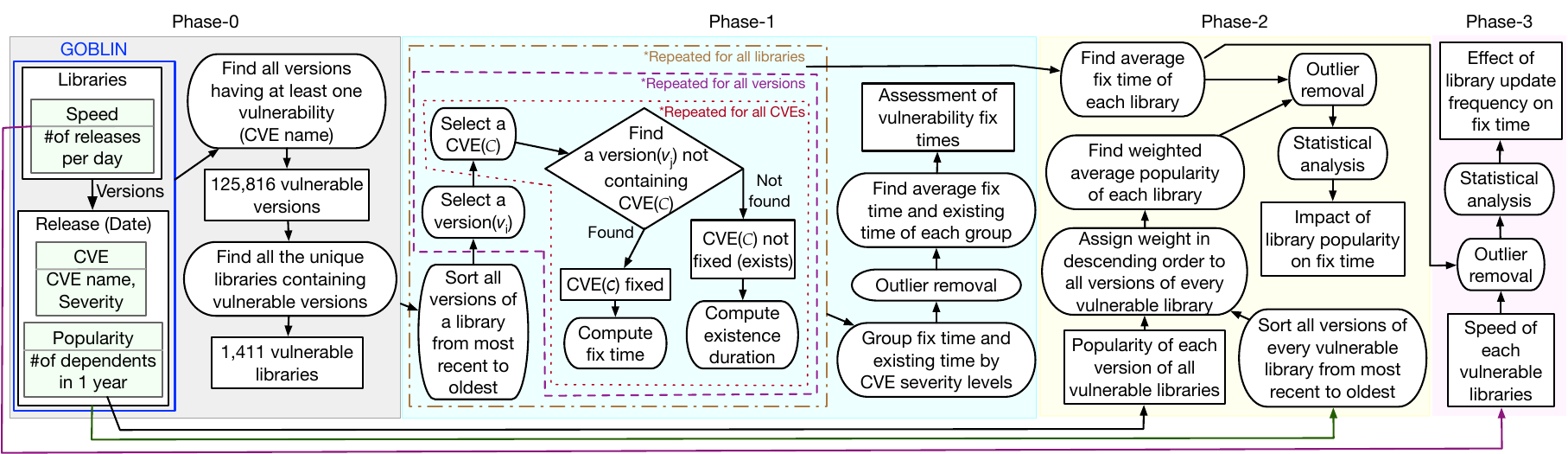}}
\vspace{-0.3cm}
\caption{Procedural steps of our study}
\vspace{-0.2cm}
\label{fig:method}
\vspace{-0.3cm}
\end{figure*}

In this study, we utilize the latest available version
of dependency graph database from the Goblin framework~\cite{jaime2024goblin}. This database contains 59,152,712 nodes which include 658,078 libraries (artifacts), 14,459,139 versions (releases), and 44,035,495 `AddedValue' nodes, which provide metrics such as \textit{CVE}, \textit{popularity}, and \textit{speed}. The set of CVEs includes information such as the \textit{name} and \textit{severity} level of each CVE for every version. The \textit{popularity} metric records the number of dependent versions for each version over a one-year period, while the \textit{speed} metric reflects the number of versions released per library each day. Libraries can have multiple versions, each associated with specific release dates.

As all three of our research questions focus on vulnerabilities, we need to extract libraries with known vulnerabilities. To achieve this, we follow the steps outlined in Phase-0 in Figure~\ref{fig:method}. If a version of a library is vulnerable, it includes a CVE name in the CVE metric. Based on this, we identify all versions with at least one vulnerability which result in a total of 125,816 vulnerable versions. Then, we identify all unique libraries containing these vulnerable versions and find a total of 1,411 vulnerable libraries.

\section{Analysis and Findings}
The steps detailed in Phase-1, Phase-2, and Phase-3 in Figure~\ref{fig:method} correspond to our methods for answering RQ-1, RQ-2, and RQ-3, respectively. 

\subsection{Severity-based Assessment of Vulnerability Fix Times}
We have already collected the release date, CVE name, and CVE severity level for all versions of each vulnerable library, as described in Section~\ref{sec:data}. To address RQ-1, we follow the steps outlined in Phase 1 of Figure~\ref{fig:method}. First, we sort all versions of each vulnerable library by their release dates, from oldest to most recent. For each version $\nu_i$ where a CVE $\mathcal{C}$ first appears, we examine subsequent versions to determine if $\mathcal{C}$ persists.

This process is repeated for all CVEs across all library versions, resulting in two outcomes:

\begin{enumerate}
    \item Detecting fix time: If we find a version $\nu_j$ where $\mathcal{C}$ is resolved, we calculate the \textit{time to resolve} as:  

\vspace{0.05cm}
{\small \emph{Time to resolve $\mathcal{C}$ = Release date of $\nu_j$ - Release date of $\nu_i$}}
\vspace{0.05cm}

    \item Identifying existence duration: If no subsequent version resolves $\mathcal{C}$, it remains unresolved as of the current date (August 30, 2024). The \textit{existence duration} is:  

\vspace{0.05cm}
{\small \emph{Existence duration of  $\mathcal{C}$ = Current Date - Release date of $\nu_i$}}

\end{enumerate}


 




After calculating the time required to fix CVEs, we group each CVE along with its fix time or existence duration according to the vulnerability severity level. Then, considering the fix time or existence duration, outliers are detected and removed using the interquartile range (IQR) method~\cite{seo2006review}. Subsequently, we compute the average fix time and existence duration for vulnerabilities within each severity group. Finally, we perform a severity-based assessment of vulnerability fix times.

\begin{table}[htbp]
    \centering
    \small
 \vspace{-0.2cm}
    \caption{Severity-based vulnerability fix times and existence duration}
    \vspace{-0.2cm}
    \label{tab:rq1}
    \setlength\tabcolsep{0.75pt}
     \begin{tabular}{|p{1.5cm}|r|r|r|@{}c@{ }|p{.5cm}|>{\raggedleft\arraybackslash}p{1.5cm}|r|}  \cline{1-4} \cline{6-8} 
\textbf{Severity}  & \multicolumn{3}{c|}{\textbf{Fix time (in days)}} & & \multicolumn{3}{c|}{\textbf{Existence duration (in days)}}\\    \cline{2-4} \cline{6-8}

&\textbf{$\mathcal{T}$*}	&\textbf{Avg.} 	&\textbf{Std. Dev.} & &\textbf{$\mathcal{T}$*}	&\textbf{Avg.} 	&\textbf{Std. Dev.}  \\ \cline{1-4} \cline{6-8}
\textbf{Critical}	&552	&\cellcolor{greenish1}{1432.59}	&1280.91	& &75	&\cellcolor{greenish1}{3844.31}	&2026.86 \\ \cline{1-4} \cline{6-8}
\textbf{High}	&1172	&1577.13	&1483.75	& &107	&4161.67	&1896.47\\ \cline{1-4} \cline{6-8}
\textbf{Moderate}	&1300	&1538.26	&1419.61	& &149	&4173.19	&1769.41\\ \cline{1-4} \cline{6-8}
\textbf{Low}	&116	&\cellcolor{pinkish}{1661.97}	&1504.48	& &10	&\cellcolor{pinkish}{4928.10}	&1090.58\\ \cline{1-4} \cline{6-8}
 \multicolumn{8}{r}{\footnotesize{*Here, $\mathcal{T}$=Total number of libraries after outlier removal}} 
   \end{tabular}
       \vspace{-0.2cm}
\end{table}

Table~\ref{tab:rq1} provides a severity-based summary of time required to fix vulnerabilities (referred to as ``Fix time") and the duration of vulnerabilities remain unresolved (referred to as ``Existence duration") for libraries, categorized by severity levels (Critical, High, Moderate, and Low). For each severity group, it includes the total number of libraries ($\mathcal{T}$) analyzed after outlier removal, the average time in days (Avg.), and the standard deviation (Std. Dev.) for both fix time and existence duration. In Table~\ref{tab:rq1}, we observe that critical vulnerabilities have the shortest average fix time (1432.59 days) among all severity levels, with comparatively lower variability (Std. Dev. = 1280.91 days). In contrast, low-severity vulnerabilities take the longest to fix, with an average of 1661.97 days, and also have the highest variability (Std. Dev. = 1504.48 days). Moderate and high vulnerabilities have average fix times of 1538.26 and 1577.13 days, respectively. This indicates that critical vulnerabilities are addressed relatively faster compared to other severity levels.

In Table~\ref{tab:rq1}, it is also observed that critical vulnerabilities exist for the shortest average time (3844.31 days), while low-severity vulnerabilities persist the longest, with an average existing time of 4928.10 days. High and Moderate vulnerabilities have similar existing times, slightly above 4150 days. This suggests that low-severity vulnerabilities tend to persist longer in libraries compared to vulnerabilities of other severity levels. 
Overall, the findings suggest that critical vulnerabilities are fixed the fastest, while low-severity vulnerabilities persist the longest.
Based on the findings, we formulate the answer to RQ-1 as follows: 
\vspace{-0.1cm}
\begin{mdframed}[style=MyFrame]
\textit{
Critical vulnerabilities are fixed faster than those of lower severity, with fix time increasing as severity decreases. 
Conversely, lower-severity vulnerabilities take longer to fix and remain unresolved for much longer. 
}
\end{mdframed} 

\subsection{Impact of Library Popularity}
To explore whether library popularity affects vulnerability fix times, we follow the steps outlined in Phase-2 of Figure~\ref{fig:method}.
First, we calculate the average time taken to address all the vulnerabilities present in each of the 1,411 vulnerable libraries. 
To examine the impact of popularity on vulnerability fix times, we need a popularity score for each library. However, the Goblin framework~\cite{jaime2024goblin} assigns popularity scores to individual versions of a library, not the library as a whole. To address this, we aggregate the popularity values of all versions of a library using a weighted average method. 
We choose this method to account for the varying significance of library versions as libraries often evolve over time, and newer versions are typically more representative of their current usage and relevance. Therefore, we assign weights to each version in descending order based on their recency, with the most recent version receiving the highest weight. This approach ensures that the popularity of newer versions is emphasized over older versions. This weighted average popularity value accurately reflects the present-day relevance of the library. 
The weighted average popularity of a library $\ell$ denoted as $P_\ell$ is computed as follows:

\vspace{-0.7cm}
\begin{equation}
P_\ell = \frac{\sum_{i\in \nu_\ell} w_i \times P^i_\ell}{\sum_{i\in \nu_\ell} w_i}
\end{equation}
\vspace{-0.4cm}

\noindent
Here, $P^i_\ell$ and $w_i$ denote the popularity and the weight of version $i$ of library $\ell$, respectively.  $\nu_\ell$ is the set of versions of library $\ell$. 
Then, to ensure more reliable analysis and prevent skewing of results, we identify and remove a total of 210 outliers using the IQR method~\cite{seo2006review}. Finally, we analyze the impact of library popularity on the time required to fix vulnerabilities.




\begin{figure}[htbp]
\vspace{-0.2cm}
\centerline{\includegraphics[scale=0.33]{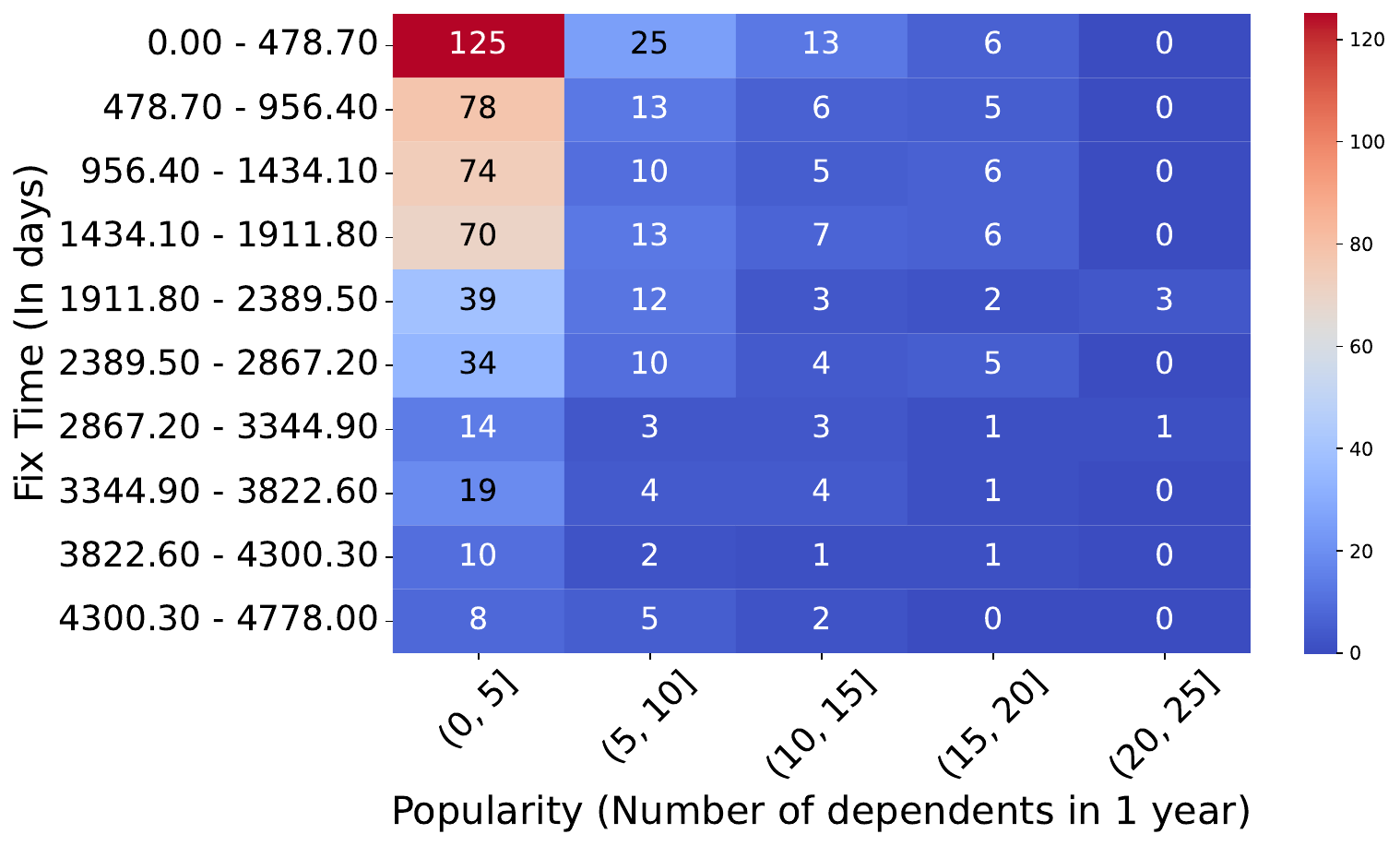}}
\vspace{-0.3cm}
\caption{Relationship between library popularity and vulnerability fix time}
\vspace{-0.1cm}
\label{fig:rq2}
\end{figure}

\begin{figure}[htbp]
\centerline{\includegraphics[scale=0.25]{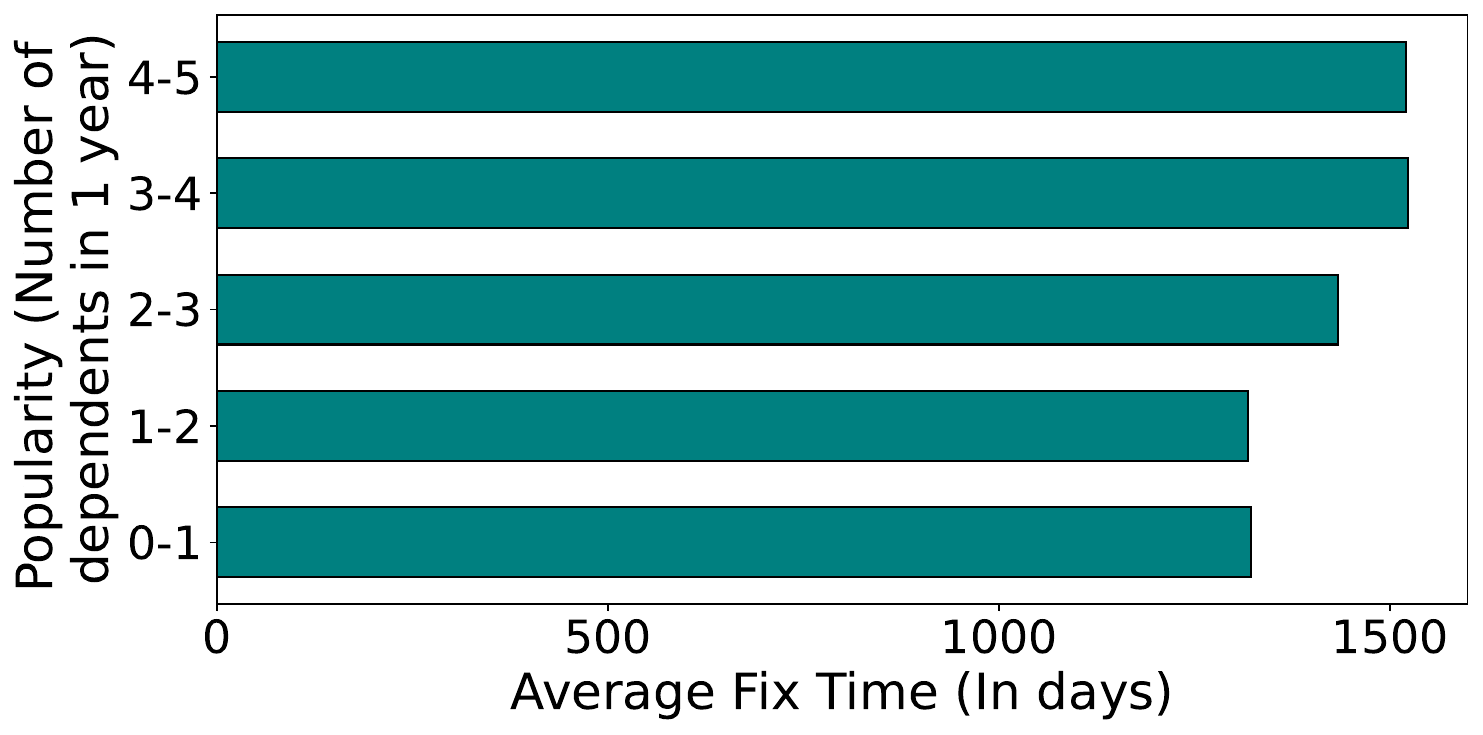}}
\vspace{-0.3cm}
\caption{Popularity and its effect on fix time (popularity scale 1-5)}
\vspace{-0.5cm}
\label{fig:rq2P}
\end{figure}

Figure~\ref{fig:rq2} presents a heatmap that illustrates the relationship between library popularity (measured by the number of dependents in one year) and the vulnerability fix times (in days). 
We group the popularity into five bins and fix-time into ten bins according to data distribution. Figure \ref{fig:rq2} suggests that majority of the instances ($\approx$72\%) have popularity score between 0 to 5, and majority of those have relatively shorter fix time as highlighted by the upper left bins. 

To better understand this relationship and confirm that our findings are statistically significant, we conduct Spearman’s correlation~\cite{szekely2023energy} and Kendall-Tau correlation~\cite{bolboaca2006pearson} tests. Both Spearman's correlation  ($\rho= 0.13,  p-value=3.81e^{-5}$) and Kendall-Tau correlation ($\rho= 0.09,  p-value=1.25e^{-17}$) tests confirm weak but significant positive correlation between popularity and fix time of vulnerabilities in libraries.  
This implies that as library popularity increases, there is a slight tendency for the average time required to fix vulnerabilities to increase.

Since the majority of the instances in Figure~\ref{fig:rq2} fall within the first popularity bin [0, 5], we further divide this bin into five equal intervals to better visualize the correlation between popularity and fix time, as our statistical tests suggest. Figure~\ref{fig:rq2P} shows the distribution of average fix time for each of these five smaller popularity bins. We observe that the average time to fix vulnerabilities increases with the increase in popularity. This suggests that less popular libraries fix vulnerabilities faster, while more popular libraries require longer times to address them.

This finding signifies the importance of looking beyond popularity while addressing vulnerabilities and regularly assessing the security of dependencies. Based on the aforementioned findings, we derive the answer to RQ-2 as follows:

\vspace{-0.2cm}
\begin{mdframed}[style=MyFrame]
\textit{
More popular libraries take slightly longer time on average to fix vulnerabilities compared to less popular libraries. 
}
\end{mdframed}

\subsection{Effect of Library Update Frequency}
To address RQ-3, we carry out the procedural steps outlined in Phase-3 in Figure~\ref{fig:method}. First, we extract the speed values for each vulnerable library, where speed represents the number of versions released per day. Libraries with high speed values are considered actively maintained, as they release new versions frequently. In contrast, libraries with low speed values are considered less actively maintained, as their version releases are infrequent. 
To ensure a more comprehensive analysis and avoid skewed results, we use the IQR method~\cite{seo2006review} to identify and remove 103 outliers based on both speed and average vulnerability fix time. Finally, we conduct statistical analysis to evaluate the effect of library speed on the time required to fix vulnerabilities.

\begin{figure}[htbp]
\centerline{\includegraphics[scale=0.33]{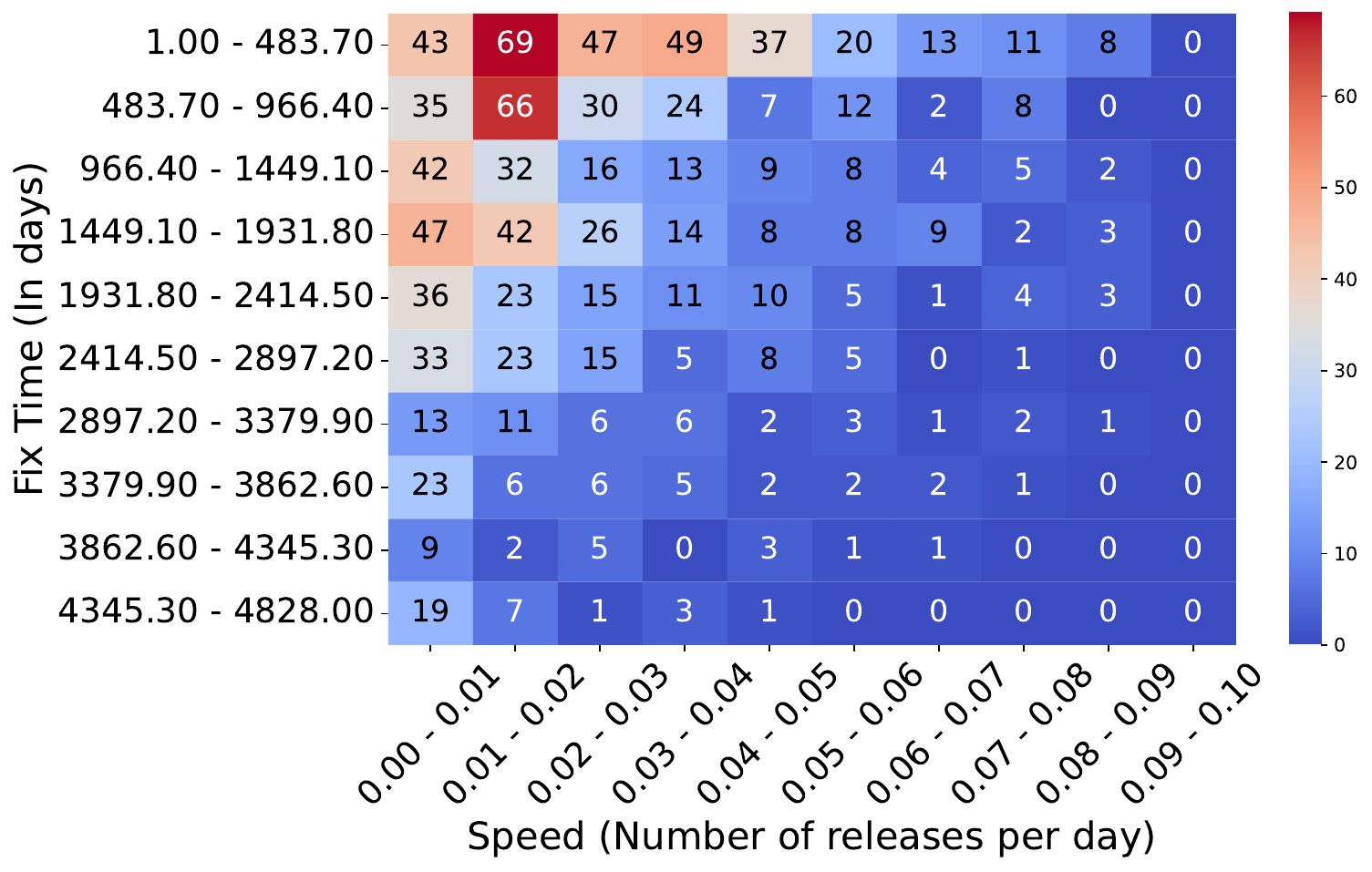}}
\vspace{-0.3cm}
\caption{Relationship between release speed and vulnerability fix time}
\vspace{-0.3cm}
\label{fig:rq3}
\end{figure}

Figure~\ref{fig:rq3} illustrates the relationship between library release speed and the time to fix vulnerabilities, measured in days. The heatmap highlights that the top-left region has the highest density, where release speed ranges between 0.00 and 0.04, and the fix time is less than 2000 days. This indicates that most projects have slower release speeds and address vulnerabilities within a moderate timeframe. In contrast, as we move from the top-left to the bottom-right, the time taken to fix vulnerabilities increases, release speeds become higher, but the number of libraries with fixed vulnerabilities decreases. Additionally, moving from left to right reveals a significant drop in the occurrence of libraries with fixed vulnerabilities across all time frames.
The findings suggest that as the speed increases, the time taken to fix vulnerabilities decreases.

To measure the strength and significance of the observed pattern, we employ Spearman’s correlation~\cite{szekely2023energy} and Kendall-Tau correlation~\cite{bolboaca2006pearson} tests. 
Both tests, Spearman's correlation  ($\rho= -0.25,  p-value=4.85e^{-17}$) and Kendall-Tau correlation ($\rho= -0.17,  p-value=1.25e^{-17}$), confirm weak but significant negative correlation between speed and fix time of vulnerabilities.
These results suggest that when speed increases (more maintenance activity), the fix time decreases (faster fixes). 
This finding highlights the critical role of active maintenance in ensuring software security and reducing the risk of prolonged exposure to vulnerabilities. 
Therefore, we attain the answer to RQ-3 as follows:

\vspace{-0.2cm}
\begin{mdframed}[style=MyFrame]
\textit{
Actively maintained libraries tend to fix vulnerabilities faster than less maintained ones, suggesting that frequent updates and releases contribute to quicker vulnerability resolutions.
}
\end{mdframed}

\section{Threats to validity}\label{sec:threats}


We focus on the Maven ecosystem, so the  insights gained from this research may not apply to all ecosystems. However, it provides valuable insights that can guide future research and be adapted for other ecosystems. 
While removing outliers reduces noise, it risks omitting crucial data that could affect the findings. To address this, we use the IQR method to ensure that the dataset remains large and reliable for analysis.

We analyze the impact of popularity (number of dependents) and speed (releases per day) separately on vulnerability resolution without considering their combined effects or other factors. Nonetheless, these well-recognized metrics allow meaningful comparisons between libraries. 
Despite these limitations, our comprehensive methodology and large dataset analysis provide reliable insights into how vulnerabilities are resolved in the Maven ecosystem.



\section{Related Work}
\vspace{-0.1cm}
Analyzing software ecosystem dependency graphs opens up research opportunities in numerous domains~\cite{Jaime2025navigating,okafor2022sok,zhang2023mitigating, cox2015measuring, bettenburg2008makes, Champa_2023_Female, Rabbi_Python_MSR2024, rabbi2024sbom, islam2024four, champa2024chatgpt, Rabbi2025chasing, Rabbi2025understanding, rabbi2025sbom}.
Bavota et al. ~\cite{bavota2013evolution} found that developers upgrade dependencies mainly for major changes.
Zerouali et al.~\cite{zerouali2022impact} revealed that many npm and RubyGems vulnerabilities stayed undisclosed for over two years and took over four years to fix.
Sabetta et al.~\cite{sabetta2024known} and Wang et al.~\cite{wang2020detecting} focused on the limitations of existing vulnerability databases and the challenge of discovering these vulnerabilities.

Jafari et al.~\cite{jafari2023dependency} revealed that younger npm packages with frequent releases and fewer dependencies address vulnerabilities faster.
Sen et al.~\cite{sen2020determinants} observed that severe vulnerabilities were disclosed quickly.
Decan et al.~\cite{decan2018impact} found delays in npm vulnerability fixes, with 15\% remaining high-risk.
Alfadel et al.~\cite{alfadel2023empirical} analyzed PyPi ecosystem and revealed that vulnerabilities often took more than three years to be identified and 40.86\% fixed post-disclosure. 

In contrast to earlier research studies that have primarily examined vulnerability within smaller or less interconnected ecosystems, our research comprehensively analyzes the Maven ecosystem using Goblin's extensive dependency graph database. We explore the impact of library popularity, version release frequency, and vulnerability severity on fix times. This study reveals how these factors impact vulnerability resolution, which is crucial for improving dependency management and mitigation efforts.

\section{Conclusion}
This research provides a comprehensive quantitative analysis of the resolution times of software vulnerabilities within the Maven system. We find that critical vulnerabilities are resolved faster than vulnerabilities of lower severity. 
Library popularity shows a weak but statistically significant positive impact on the speed of vulnerability resolution. This indicates that as the number of dependents increases, vulnerabilities are addressed more quickly.
Libraries with frequent version releases, demonstrate faster vulnerability fixes, highlighting the importance of active maintenance.
The findings are derived from a detailed examination of over 4,459,139 versions across 658,078 libraries using the dependency graph database of the Goblin framework. The findings are statistically significant and provide valuable insights for improving vulnerability management strategies in open-source ecosystems. Future work will address the limitations identified in this study, such as exploring the combined effects of popularity and release frequency and incorporating additional metrics to better understand the factors influencing vulnerability resolution. 

\balance

\section*{Acknowledgement}
This work is supported in part by the ISU-CAES (Center for Advanced Energy Studies) Seed Grant at the Idaho State University, USA.

\bibliographystyle{IEEEtran}
\bibliography{MSR25AIC, gender, DevGPT_Task}

\begin{thebibliography}{10}
\providecommand{\url}[1]{#1}
\csname url@samestyle\endcsname
\providecommand{\newblock}{\relax}
\providecommand{\bibinfo}[2]{#2}
\providecommand{\BIBentrySTDinterwordspacing}{\spaceskip=0pt\relax}
\providecommand{\BIBentryALTinterwordstretchfactor}{4}
\providecommand{\BIBentryALTinterwordspacing}{\spaceskip=\fontdimen2\font plus
\BIBentryALTinterwordstretchfactor\fontdimen3\font minus
  \fontdimen4\font\relax}
\providecommand{\BIBforeignlanguage}[2]{{%
\expandafter\ifx\csname l@#1\endcsname\relax
\typeout{** WARNING: IEEEtran.bst: No hyphenation pattern has been}%
\typeout{** loaded for the language `#1'. Using the pattern for}%
\typeout{** the default language instead.}%
\else
\language=\csname l@#1\endcsname
\fi
#2}}
\providecommand{\BIBdecl}{\relax}
\BIBdecl

\bibitem{zhang2020companies}
Y.~Zhang, M.~Zhou, K.-J. Stol, J.~Wu, and Z.~Jin, ``How do companies
  collaborate in open source ecosystems? an empirical study of openstack,'' in
  \emph{Proceedings of the ACM/IEEE 42nd International Conference on Software
  Engineering (ICSE)}, 2020, pp. 1196--1208.

\bibitem{meszaros2024dynamics}
G.~M{\'e}sz{\'a}ros and J.~Wachs, ``The dynamics of innovation in open source
  software ecosystems,'' \emph{arXiv preprint arXiv:2411.14894}, 2024.

\bibitem{bals2024open}
\BIBentryALTinterwordspacing
F.~Bals, ``2024 open source security and risk analysis report,'' Feb 2024,
  accessed: 2024-11-27. [Online]. Available:
  \url{https://www.blackduck.com/blog/open-source-trends-ossra-report.html}
\BIBentrySTDinterwordspacing

\bibitem{alfadel2023empirical}
M.~Alfadel, D.~E. Costa, and E.~Shihab, ``Empirical analysis of security
  vulnerabilities in python packages,'' \emph{Empirical Software Engineering},
  vol.~28, no.~3, p.~59, 2023.

\bibitem{wang2023deepvd}
W.~Wang, T.~N. Nguyen, S.~Wang, Y.~Li, J.~Zhang, and A.~Yadavally, ``Deepvd:
  Toward class-separation features for neural network vulnerability
  detection,'' in \emph{2023 IEEE/ACM 45th International Conference on Software
  Engineering (ICSE)}.\hskip 1em plus 0.5em minus 0.4em\relax IEEE, 2023, pp.
  2249--2261.

\bibitem{zhang2023mitigating}
L.~Zhang, C.~Liu, S.~Chen, Z.~Xu, L.~Fan, L.~Zhao, Y.~Zhang, and Y.~Liu,
  ``Mitigating persistence of open-source vulnerabilities in maven ecosystem,''
  in \emph{2023 38th IEEE/ACM International Conference on Automated Software
  Engineering (ASE)}.\hskip 1em plus 0.5em minus 0.4em\relax IEEE, 2023, pp.
  191--203.

\bibitem{okafor2022sok}
C.~Okafor, T.~R. Schorlemmer, S.~Torres-Arias, and J.~C. Davis, ``Sok: Analysis
  of software supply chain security by establishing secure design properties,''
  in \emph{Proc. of the 2022 ACM Workshop on Software Supply Chain Offensive
  Research and Ecosystem Defenses}, 2022, pp. 15--24.

\bibitem{ptsecurity2024consequences}
\BIBentryALTinterwordspacing
{PT Security}, ``The consequences of delays in remediating vulnerabilities,
  2022–2023,'' Jul 2024, accessed: 2024-11-27. [Online]. Available:
  \url{https://global.ptsecurity.com/analytics/the-consequences-of-delays-in-remediating-vulnerabilities-2022-2023}
\BIBentrySTDinterwordspacing

\bibitem{sen2020determinants}
R.~Sen, J.~Choobineh, and S.~Kumar, ``Determinants of software vulnerability
  disclosure timing,'' \emph{Production and Operations Management}, vol.~29,
  no.~11, pp. 2532--2552, 2020.

\bibitem{venkat2023unpatched}
\BIBentryALTinterwordspacing
A.~Venkat, ``Unpatched old vulnerabilities continue to be exploited: Report,''
  Mar 2023, accessed: 2024-11-27. [Online]. Available:
  \url{https://www.csoonline.com/article/574661/unpatched-old-vulnerabilities-continue-to-be-exploited-report.html}
\BIBentrySTDinterwordspacing

\bibitem{brewer2021know}
\BIBentryALTinterwordspacing
E.~Brewer, R.~Pike, A.~Arya, A.~Bertucio, and K.~Lewandowski, ``Know, prevent,
  fix: A framework for shifting the discussion around vulnerabilities in open
  source,'' Feb 2021, accessed: 2024-11-27. [Online]. Available:
  \url{https://opensource.googleblog.com/2021/02/know-prevent-fix-framework-for-shifting-discussion-around-vulnerabilities-in-open-source.html}
\BIBentrySTDinterwordspacing

\bibitem{weigel2024silent}
\BIBentryALTinterwordspacing
B.~H. Weigel, ``The silent crisis in open source: When maintainers walk away,''
  Jul 2024, accessed: 2024-11-27. [Online]. Available:
  \url{https://opensauced.pizza/blog/when-open-source-maintainers-leave}
\BIBentrySTDinterwordspacing

\bibitem{Jaime2025navigating}
D.~Jaime, J.~El~Haddad, and P.~Poizat, ``Navigating and exploring software
  dependency graphs using goblin,'' in \emph{Proceedings of the International
  Conference on Mining Software Repositories (MSR)}, 2025.

\bibitem{replicationPackage}
M.~Rabbi, A.~Champa, R.~Paul, and M.~Zibran, ``Vulnerablility assessment in
  maven: Replication package,''
  \url{https://doi.org/10.6084/m9.figshare.27984830.v1}, December 2024.

\bibitem{jaime2024goblin}
D.~Jaime, ``Goblin: Neo4j maven central dependency graph,''
  \url{https://zenodo.org/records/13734581}, August 2024.

\bibitem{seo2006review}
S.~Seo, ``A review and comparison of methods for detecting outliers in
  univariate data sets,'' Ph.D. dissertation, University of Pittsburgh, 2006.

\bibitem{szekely2023energy}
G.~J. Sz{\'e}kely and M.~L. Rizzo, \emph{The energy of data and distance
  correlation}.\hskip 1em plus 0.5em minus 0.4em\relax CRC Press, 2023.

\bibitem{bolboaca2006pearson}
S.-D. Bolboaca and L.~J{\"a}ntschi, ``Pearson versus spearman, kendall’s tau
  correlation analysis on structure-activity relationships of biologic active
  compounds,'' \emph{Leonardo J. of Sciences}, vol.~5, no.~9, pp. 179--200,
  2006.

\bibitem{cox2015measuring}
J.~Cox, E.~Bouwers, M.~Van~Eekelen, and J.~Visser, ``Measuring dependency
  freshness in software systems,'' in \emph{2015 IEEE/ACM 37th IEEE
  International Conference on Software Engineering}, vol.~2.\hskip 1em plus
  0.5em minus 0.4em\relax IEEE, 2015, pp. 109--118.

\bibitem{bettenburg2008makes}
N.~Bettenburg, S.~Just, A.~Schr{\"o}ter, C.~Weiss, R.~Premraj, and
  T.~Zimmermann, ``What makes a good bug report?'' in \emph{Proceedings of the
  16th ACM SIGSOFT International Symposium on Foundations of software
  engineering}, 2008, pp. 308--318.

\bibitem{Champa_2023_Female}
A.~Champa, M.~Rabbi, M.~Zibran, and M.~Islam, ``Insights into female
  contributions in open-source projects,'' in \emph{20th IEEE International
  Conference on Mining Software Repositories}, 2023, pp. 357--361.

\bibitem{Rabbi_Python_MSR2024}
M.~F. Rabbi, A.~I. Champa, M.~F. Zibran, and M.~R. Islam, ``{AI} writes, we
  analyze: The {ChatGPT} python code saga,'' in \emph{Proceedings of ACM
  International Conference on Mining Software Repositories (MSR 2024)}, 2024.

\bibitem{rabbi2024sbom}
M.~F. Rabbi, A.~I. Champa, C.~Nachuma, and M.~F. Zibran, ``Sbom generation
  tools under microscope: A focus on the npm ecosystem,'' in \emph{Proceedings
  of the 39th ACM/SIGAPP Symposium on Applied Computing}, 2024, pp. 1233--1241.

\bibitem{islam2024four}
M.~R. Islam, M.~F. Rabbi, J.~Youngeun, A.~I. Champa, E.~Young, C.~Wilson, G.~J.
  Scott, and M.~F. Zibran, ``A four-dimension gold standard dataset for opinion
  mining in software engineering,'' in \emph{Proceedings of the 21st
  International Conference on Mining Software Repositories}, 2024, pp.
  487--491.

\bibitem{champa2024chatgpt}
A.~I. Champa, M.~F. Rabbi, C.~Nachuma, and M.~F. Zibran, ``Chatgpt in action:
  Analyzing its use in software development,'' in \emph{Proceedings of the 21st
  International Conference on Mining Software Repositories}, 2024, pp.
  182--186.

\bibitem{Rabbi2025chasing}
M.~F. Rabbi, A.~I. Champa, R.~Paul, and M.~F. Zibran, ``Chasing the clock: How
  fast are vulnerabilities fixed in the maven ecosystem?'' in \emph{Proceedings
  of the 22nd International Conference on Mining Software Repositories}, 2024,
  pp. 1--5 (to appear).

\bibitem{Rabbi2025understanding}
M.~F. Rabbi, R.~Paul, A.~I. Champa, and M.~F. Zibran, ``Understanding software
  vulnerabilities in the maven ecosystem: Patterns, timelines, and risks,'' in
  \emph{Proceedings of the 22nd International Conference on Mining Software
  Repositories}, 2024, pp. 1--5 (to appear).

\bibitem{rabbi2025sbom}
M.~F. Rabbi, A.~I. Champa, and M.~F. Zibran, ``Sbom generation tools under
  microscope: A focus on the npm ecosystem,'' in \emph{Proceedings of the 40th
  ACM/SIGAPP Symposium on Applied Computing}, 2025, pp. 1--9 (to appear).

\bibitem{bavota2013evolution}
G.~Bavota, G.~Canfora, M.~Di~Penta, R.~Oliveto, and S.~Panichella, ``The
  evolution of project inter-dependencies in a software ecosystem: The case of
  apache,'' in \emph{2013 IEEE international conference on software
  maintenance}.\hskip 1em plus 0.5em minus 0.4em\relax IEEE, 2013, pp.
  280--289.

\bibitem{zerouali2022impact}
A.~Zerouali, T.~Mens, A.~Decan, and C.~De~Roover, ``On the impact of security
  vulnerabilities in the npm and rubygems dependency networks,''
  \emph{Empirical Software Engineering}, vol.~27, no.~5, p. 107, 2022.

\bibitem{sabetta2024known}
A.~Sabetta, S.~E. Ponta, R.~C. Lozoya, M.~Bezzi, T.~Sacchetti, M.~Greco,
  G.~Balogh, P.~Heged{\H{u}}s, R.~Ferenc, R.~Paramitha \emph{et~al.}, ``Known
  vulnerabilities of open source projects: Where are the fixes?'' \emph{IEEE
  Security \& Privacy}, 2024.

\bibitem{wang2020detecting}
W.~Wang, F.~Dumont, N.~Niu, and G.~Horton, ``Detecting software security
  vulnerabilities via requirements dependency analysis,'' \emph{IEEE
  Transactions on Soft. Engineering}, vol.~48, no.~5, pp. 1665--1675, 2020.

\bibitem{jafari2023dependency}
A.~Jafari, D.~Costa, A.~Abdellatif, and E.~Shihab, ``Dependency practices for
  vulnerability mitigation,'' \emph{arXiv preprint arXiv:2310.07847}, 2023.

\bibitem{decan2018impact}
A.~Decan, T.~Mens, and E.~Constantinou, ``On the impact of security
  vulnerabilities in the npm package dependency network,'' in \emph{Proceedings
  of the 15th international conference on mining software repositories}, 2018,
  pp. 181--191.

\end{thebibliography}
\end{document}